\documentclass[12pt,a4paper]{article}
\usepackage{amssymb}
\usepackage{graphicx,color}
\makeatletter
\def\rddots{\mathinner{\mkern1mu\raise\p@%
     \vbox{\kern7\p@\hbox{.}}\mkern2mu%
     \raise4\p@\hbox{.}\mkern2mu\raise7\p@\hbox{.}\mkern1mu}}
\makeatother
\makeatletter
\def\eqnarray{%
\stepcounter{equation}%
\let\@currentlabel=\theequation
\global\@eqnswtrue
\global\@eqcnt\z@
\tabskip\@centering
\let\\=\@eqncr
$$\halign to \displaywidth\bgroup\@eqnsel\hskip\@centering
$\displaystyle\tabskip\z@{##}$&\global\@eqcnt\@ne
\hfil$\displaystyle{{}##{}}$\hfil
&\global\@eqcnt\tw@$\displaystyle\tabskip\z@{##}$\hfil
\tabskip\@centering&\llap{##}\tabskip\z@\cr}
\makeatother
\newcommand{\ket}[1]{{\vert{#1}\rangle}}

\newcommand{\fukuso}{{\mathbf C}}
\newcommand{\real}{{\mathbf R}}
\newcommand{\futon}{{\bf N}}

\begin{document}

\title{\sl Exponentiation of certain Matrices related to the Four 
Level System by use of
the Magic Matrix}

%
\author{
   Kazuyuki FUJII
   \thanks{E-mail address : fujii@yokohama-cu.ac.jp }\\
   Department of Mathematical Sciences\\
   Yokohama City University\\
   Yokohama, 236--0027\\
   Japan
   }
\date{6 March 2007}
\maketitle
%
%
%
%
\begin{abstract}

   In this paper we show  how to calculate explicitly the exponential 
of certain matrices,
  which are evolution operators governing the interaction of the four 
level system of atoms and the radiation, etc.  We present a consistent
   method in terms of the magic matrix by Makhlin.

As a closely related subject, we derive a closed form expression of the
Baker-Campbell-Hausdorff  formula for a class
   of matrices in $SU(4)$,  by use of the method developed by the 
present authors in quant-ph/0610009.

%

\end{abstract}
%


%
%
%
%

\section{Introduction}

The purpose of this paper is to develop a useful method to calculate the
exponential of certain matrices explicitly.
These matrices arise from differential equations governing the 
interaction of  the
four level system of atoms with the radiation whose image is 
illustrated in Figure 1.

\vspace{5mm}
\begin{center}
  \input{atom-with-4-energy-levels.fig}
\end{center}
\begin{center}
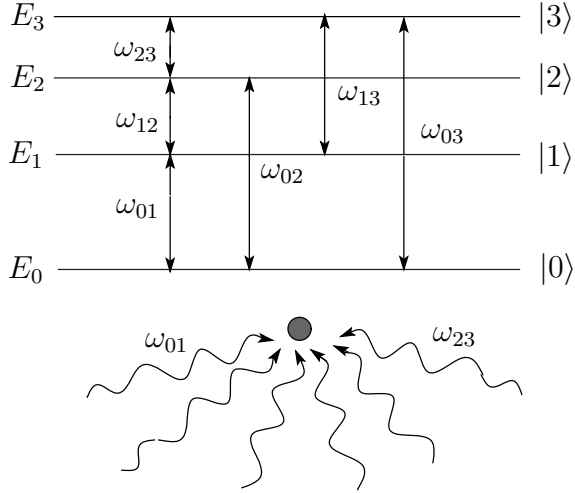
  Fig.1\ Atom with four energy levels and general action by laser fields
\end{center}

\par \noindent
For the general background, motivation and possible applications of 
the present work,  see the recent work of the author and 
collaborators \cite{FHKW}, \cite{KF1} and
\cite{FF}. Under the RWA (\underline{R}otating \underline{W}ave
\underline{A}pproximation) and some resonance conditions
the problem is reduced to the evaluation of  the exponential (i.e. 
finite time evolution operator) $\mbox{e}^{-itH}$
  with the Hamiltonian
\[
H=
\left(
   \begin{array}{cccc}
     0 & h_{12} & h_{13} & h_{14}  \\
     h_{12} & 0 & h_{23} & h_{24}  \\
     h_{13} & h_{23} & 0 & h_{34}  \\
     h_{14} & h_{24} & h_{34} & 0
   \end{array}
\right),
\]
where $h_{ij}$ are real coupling constants between the atom and laser fields.
Here we have changed some notation from the previous one in 
\cite{KF1} (like $g_{ij}\leftrightarrow
h_{i-1,j-1}$) for convenience.
A generic algorithm to calculate $\mbox{e}^{-itH}$ based on eigenvalues of the
Hamiltonian $H$ has been given by us in \cite{KF1}, although the 
actual execution is not so simple.

In this paper we revisit the problem from a different point of view.
Let us introduce  Makhlin's theorem. The isomorphism
\[
SU(2)\otimes SU(2)\cong SO(4)\quad (\Longleftrightarrow
su(2)\otimes 1_{2}+1_{2}\otimes su(2) \cong so(4))
\]
is one of the well-known theorems in elementary representation theory and is
characteristic of four dimensional Euclidean space.
In \cite{YMa} Makhlin gave it the adjoint expression explicitly:
\[
F : SU(2)\otimes SU(2) \longrightarrow SO(4),\quad
F(A\otimes B)=Q^{\dagger}(A\otimes B)Q,
\]
with some unitary matrix $Q\in U(4)$. As far as we know this is the 
first time that
the map was realized by the adjoint action. See also \cite{FOS}, where
a slightly different matrix $R$ has been used in place of $Q$.
This $R$ (in our notation) is interesting enough and is called the magic
matrix by Makhlin, see also \cite{ZVWS} and \cite{RZ}.

As the Hamiltonian $H$ above is real symmetric ($\in su(4)$) (not 
anti--symmetric
($\in so(4)$)) it is not obvious  whether
the magic matrix $R$ could be applied to it or not.
As we will show presently, a fairly wide sub-class of such 
Hamiltonians can be explicitly
exponentiated with the help of the magic matrix and an additional 
similarity transformation. For the generic Hamiltonian $H$, we 
provide an approximate result of explicit exponentiation.

Here we note that there is some overlap between our work and \cite{RZ}.
However, the methods given  in \cite{RZ} are quite varied and rather 
complicated.
On the other hand ours rely on a consistent use of the magic
matrix, so we believe our method could be easily digested  by general 
readers, though the scope might be limited.

\vspace{3mm}
As a closely related subject, we discuss the Baker-Campbell-Hausdorff 
(B-C-H) formula. It is one of
the fundamental theorems in elementary Linear Algebra (or Lie group):
\[
\mbox{e}^{A}\mbox{e}^{B}=\mbox{e}^{{BCH}(A,B)}\ ;\
{BCH}(A,B)=A+B+\frac{1}{2}[A,B]+
\frac{1}{12}\left\{[[A,B],B]+[A,[A,B]]\right\}+\cdots,
\]
where $A$ and $B$ are elements of some algebra. See for example the textbooks
\cite{Vara}, \cite{YS} or \cite{SW}.
A closed expression for the B-C-H formula for $SU(2)$
is quite well-known.  Similar closed expression  is obtained
for  $SO(4)$ (namely, for $A, B \in so(4)$) by making use of the magic matrix,
see \cite{KF1}.

Here we address the problem of explicit summation of the right hand
side of the B-C-H formula for certain lower dimensional matrices, in 
particular, those
related to $su(4)$.


In this paper we treat matrices of type
\[
H=
\left(
   \begin{array}{cccc}
     0 & h_{12} & 0 & h_{14}       \\
     h_{12} & 0 & h_{23} & 0  \\
     0 & h_{23} & 0 & h_{34}  \\
     h_{14} & 0 & h_{34} & 0
   \end{array}
\right),
\]
%
which are very important in quantum optics, quantum computation, etc. 
(see \cite{FHKW}). We will derive  its exact exponential form 
$\mbox{e}^{-itH}$ and will present   a closed form expression of the 
B-C-H formula
for  two of them.

\section{Magic Matrix}

In this section we introduce appropriate  concepts and notation 
together with a brief review
of  the results in \cite{FOS} within our necessity, which look
slightly different from the original work of Makhlin  in \cite{YMa}.

The 1-qubit space is $\fukuso^{2}=\mbox{Vect}_{\fukuso}\{\ket{0},\ket{1}\}$
where
\begin{equation}
\label{eq:bra-ket}
\ket{0}=
\left(
\begin{array}{c}
  1 \\
  0
\end{array}
\right),
\quad
\ket{1}=
\left(
\begin{array}{c}
  0 \\
  1
\end{array}
\right).
\end{equation}
Let $\{\sigma_{1}, \sigma_{2}, \sigma_{3}\}$ be the Pauli matrices acting on
the space
\begin{equation}
\label{eq:Pauli matrices}
\sigma_{1} =
\left(
   \begin{array}{cc}
     0 & 1 \\
     1 & 0
   \end{array}
\right), \quad
\sigma_{2} =
\left(
   \begin{array}{cc}
     0 & -i \\
     i & 0
   \end{array}
\right), \quad
\sigma_{3} =
\left(
   \begin{array}{cc}
     1 & 0 \\
     0 & -1
   \end{array}
\right)
\quad \mbox{and} \quad
1_{2} =
\left(
   \begin{array}{cc}
     1 & 0 \\
     0 & 1
   \end{array}
\right).
\end{equation}

Next let us consider the 2--qubit space.
Now we use the tensor product  notation which is different from the usual one.
That is,
\[
\fukuso^{2}{\otimes}\fukuso^{2}=\{a\otimes b\ |\ a,b\in \fukuso^{2}\},\
\fukuso^{2}\widehat{\otimes}\fukuso^{2}=
\left\{\sum_{j=1}^{k}\lambda_{j}a_{j}\otimes b_{j}\ |\ a_{j},b_{j}\in
\fukuso^{2},\ \lambda_{j}\in \fukuso,\ k\in \futon \right\}\cong \fukuso^{4}.
\]
Then the 2-qubit space is
\[
\fukuso^{2}\widehat{\otimes}\fukuso^{2}=\mbox{Vect}_{\fukuso}
\{\ket{00},\ket{01},\ket{10},\ket{11}\},
\]
where $\ket{ab}=\ket{a}\otimes \ket{b}\ (a,b\in \{0,1\})$.
The Bell basis
$\{\ket{\Psi_{1}},\ket{\Psi_{2}},\ket{\Psi_{3}},\ket{\Psi_{4}}\}$
defined by
{\small
\begin{equation}
\label{eq:Bell bases}
\ket{\Psi_{1}}=\frac{1}{\sqrt{2}}(\ket{00}+\ket{11}),\
\ket{\Psi_{2}}=\frac{1}{\sqrt{2}}(\ket{01}+\ket{10}),\
\ket{\Psi_{3}}=\frac{1}{\sqrt{2}}(\ket{01}-\ket{10}),\
\ket{\Psi_{4}}=\frac{1}{\sqrt{2}}(\ket{00}-\ket{11})
\end{equation}
}
plays an important role in various context.

By $H_{0}(2;\fukuso)$ we denote the set of all traceless hermitian matrices in
$M(2;\fukuso)$. It is well-known that they are spanned by the Pauli matrices
\[
H_{0}(2;\fukuso)=
\{{\bf a}\equiv a_{1}\sigma_{1}+a_{2}\sigma_{2}+a_{3}\sigma_{3}\
|\ a_{1},a_{2},a_{3}\in \real\}
\]
and $H_{0}(2;\fukuso)\cong su(2)$ where $su(2)=\mathfrak{L}(SU(2))$ is
the Lie algebra of the group $SU(2)$.

\bigskip

The isomorphism is realized as the adjoint action (the Makhlin's theorem)
as follows
\[
F : SU(2)\otimes SU(2) \longrightarrow SO(4),\quad
F(A\otimes B)=R^{\dagger}(A\otimes B)R
\]
where
\begin{equation}
R\equiv
\left(
\ket{\Psi_{1}},-i\ket{\Psi_{2}},-\ket{\Psi_{3}},-i\ket{\Psi_{4}}
\right)
=\frac{1}{\sqrt{2}}
\left(
   \begin{array}{cccc}
     1 &  0 &  0 & -i  \\
     0 & -i & -1 &  0  \\
     0 & -i &  1 &  0  \\
     1 &  0 &  0 &  i
   \end{array}
\right).
\end{equation}
Note that the unitary matrix $R$ is a bit different from $Q$ in \cite{YMa}.

\vspace{3mm}
Let us consider this isomorphism at the Lie algebra level because it 
is in general
  easier than  at the  Lie group level:
\vspace{5mm}
\begin{center}
\input{Lie-diagram.fig}
\end{center}

\vspace{5mm} \noindent
Since the Lie algebra of $SU(2)\otimes SU(2)$ is
\[
\mathfrak{L}(SU(2)\otimes SU(2))=
su(2)\otimes 1_{2}+1_{2}\otimes su(2)=
\left\{i({\bf a}\otimes 1_{2}+1_{2}\otimes {\bf b})\ |\ {\bf a},\ {\bf b} \in
H_{0}(2;\fukuso)\right\},
\]
we have only to examine
\begin{equation}
\label{eq:Lie algebra level}
f(i({\bf a}\otimes 1_{2}+1_{2}\otimes {\bf b}))=
iR^{\dagger}({\bf a}\otimes 1_{2}+1_{2}\otimes {\bf b})R\in
\mathfrak{L}(SO(4))\equiv so(4).
\end{equation}
If we set ${\bf a}=\sum_{j=1}^{3}a_{j}\sigma_{j}$ and
${\bf b}=\sum_{j=1}^{3}b_{j}\sigma_{j}$
then the right hand side of (\ref{eq:Lie algebra level}) reads
\begin{equation}
\label{eq:correspondence}
iR^{\dagger}({\bf a}\otimes 1_{2}+1_{2}\otimes {\bf b})R
=
\left(
   \begin{array}{cccc}
     0 & a_{1}+b_{1} & a_{2}-b_{2} & a_{3}+b_{3}       \\
     -(a_{1}+b_{1}) & 0 & a_{3}-b_{3} & -(a_{2}+b_{2}) \\
     -(a_{2}-b_{2}) & -(a_{3}-b_{3}) & 0 & a_{1}-b_{1} \\
     -(a_{3}+b_{3}) & a_{2}+b_{2} & -(a_{1}-b_{1}) & 0
   \end{array}
\right).
\end{equation}
Conversely, if
\[
A=
\left(
   \begin{array}{cccc}
     0 & f_{12} & f_{13} & f_{14}   \\
    -f_{12} & 0 & f_{23} & f_{24}   \\
    -f_{13} & -f_{23} & 0 & f_{34}  \\
    -f_{14} & -f_{24} & -f_{34} & 0
   \end{array}
\right)\in so(4)
\]
then we obtain
\begin{equation}
\label{eq:normal-adjoint}
RAR^{\dagger}=i({\bf a}\otimes 1_{2}+1_{2}\otimes {\bf b})
\end{equation}
with
\begin{eqnarray}
\label{eq:left-a}
&&{\bf a}=a_{1}\sigma_{1}+a_{2}\sigma_{2}+a_{3}\sigma_{3}
    =\frac{f_{12}+f_{34}}{2}\sigma_{1}+
     \frac{f_{13}-f_{24}}{2}\sigma_{2}+
     \frac{f_{14}+f_{23}}{2}\sigma_{3}, \\
\label{eq:right-b}
&&{\bf b}=b_{1}\sigma_{1}+b_{2}\sigma_{2}+b_{3}\sigma_{3}
    =\frac{f_{12}-f_{34}}{2}\sigma_{1}-
     \frac{f_{13}+f_{24}}{2}\sigma_{2}+
     \frac{f_{14}-f_{23}}{2}\sigma_{3}.
\end{eqnarray}
It is very interesting to note that ${\bf a}$ and ${\bf b}$ are the 
self-dual  and
anti-self-dual part of the matrix $A$, respectively, under the Hodge 
$*$-operation
defined by $(*F)_{ij}=\frac{1}{2}\sum_{k,l=1}^4\epsilon_{ijkl}F_{kl}$.
Here $\epsilon_{ijkl}$ is the totally anti-symmetric tensor with 
$\epsilon_{1234}=1$.

\section{The Exponential of Matrices in Four Level System}

This section provides the main results of the paper,  explicit calculation of
the exponential of certain matrices.
The generic  Hamiltonian  treated in this paper is a real symmetric 
$4\times 4$ matrix
with vanishing diagonal elements
\begin{equation}
\label{eq:hamiltonian}
H=
\left(
   \begin{array}{cccc}
     0 & h_{12} & h_{13} & h_{14}  \\
     h_{12} & 0 & h_{23} & h_{24}  \\
     h_{13} & h_{23} & 0 & h_{34}  \\
     h_{14} & h_{24} & h_{34} & 0
   \end{array}
\right),\qquad h_{ij}\in\bf{R}.
\end{equation}
The general method is quite simple:
evaluate the evolution operator (matrix) $U(t)\equiv
\mbox{e}^{-itH}$ by unitary conjugation of the Hamiltonian $H$  in terms of the
magic matrix $R$:
\begin{equation}
\label{eq:formula}
U(t)=\mbox{e}^{-itH}
=RR^{\dagger}\mbox{e}^{-itH}RR^{\dagger}
=R\mbox{e}^{-itR^{\dagger}HR}R^{\dagger}.
\end{equation}
The conjugated Hamiltonian $R^{\dagger}HR$ reads explicitly
{\footnotesize
\begin{eqnarray}
\label{eq:}
&&R^{\dagger}HR=\frac{1}{2}\times  \nonumber \\
&&\left(
   \begin{array}{cccc}
     2h_{14} & -i(h_{12}+h_{13}+h_{24}+h_{34}) & -h_{12}+h_{13}-h_{24}+h_{34}
     & 0  \\
     i(h_{12}+h_{13}+h_{24}+h_{34}) & 2h_{23} & 0 &
     h_{12}+h_{13}-h_{24}-h_{34} \\
    -h_{12}+h_{13}-h_{24}+h_{34} & 0 & -2h_{23} &
     i(h_{12}-h_{13}-h_{24}+h_{34})  \\
     0 & h_{12}+h_{13}-h_{24}-h_{34} & -i(h_{12}-h_{13}-h_{24}+h_{34})
      & -2h_{14}
   \end{array}
\right),   \nonumber \\
&&{}
\end{eqnarray}
}
%
whose structure is better displayed in the following tensor product notation
\begin{eqnarray}
\label{eq:adjoint-form}
R^{\dagger}HR=&&
\left(\frac{h_{13}-h_{24}}{2}\sigma_{1}+
\frac{h_{14}+h_{23}}{2}\sigma_{3}\right)\otimes 1_{2}
+
1_{2}\otimes \left(\frac{h_{13}+h_{24}}{2}\sigma_{2}+
\frac{h_{14}-h_{23}}{2}\sigma_{3}\right)       \nonumber \\
&&+ \frac{h_{12}+h_{34}}{2}\sigma_{3}\otimes \sigma_{2}-
\frac{h_{12}-h_{34}}{2}\sigma_{1}\otimes \sigma_{3}.
\end{eqnarray}
Obviously the first two terms commute with each other and  the last two terms
$\sigma_{3}\otimes \sigma_{2}$ and $\sigma_{1}\otimes \sigma_{3}$
also commute.  The right hand side of (\ref{eq:adjoint-form}) is a summation of
two blocks not commuting each other.


\bigskip
A few comments  are in order.

\par \noindent
(1)\ In the calculation above we have used $R^{\dagger}HR$ in stead of
$RHR^{\dagger}$ in (\ref{eq:normal-adjoint}). This point is important (we
leave it to the readers to contemplate why it is so).

\par \noindent
(2)\ The space of the entangled states in the two-qubit system is
$SU(4)/SU(2)\otimes SU(2)$, which is identified with the homogeneous 
space $SU(4)/SO(4)$
because of the isomorphism $SU(2)\otimes SU(2)\cong SO(4)$.
Then it is well-known that
\[
\left\{A\in SU(4)\ |\ A^{T}=A\right\}
      =\left\{A^{T}A\ |\ A\in SU(4)\right\}
      \cong SU(4)/SO(4),
\]
see for example \cite{FOS}. By use of the expression
$SU(4)\ni A=\mbox{e}^{iK}$ with $K\in H_{0}(4;\fukuso)$ we have
\[
SU(4)/SO(4)\cong \left\{\mbox{e}^{iK}\ |\ K \in H_{0}(4;\real)\right\},
\]
where $H_{0}(4;\real)$ is the set of all traceless symmetric matrices in
$M(4;\real)$. It is of dimension 9 (see (\ref{Kform})), as expected: 
$15 (su(4))- 6 (so(4))=9$.

Since this $K$ is written as
\begin{equation}
K=
\left(
   \begin{array}{cccc}
     h_{1} & h_{12} & h_{13} & h_{14}  \\
     h_{12} & h_{2} & h_{23} & h_{24}  \\
     h_{13} & h_{23} & h_{3} & h_{34}  \\
     h_{14} & h_{24} & h_{34} & h_{4}
   \end{array}
\right)
=\mbox{diag}(h_{1},h_{2},h_{3},h_{4})+H,\quad h_{1}+h_{2}+h_{3}+h_{4}=0,
\label{Kform}
\end{equation}
%
%
the evaluation of the evolution operator $U(t)=\mbox{e}^{-itH}$ is 
deeply related with the study of
the entangled states in the two-qubit systems.

For later convenience  the conjugated matrix of $K$ is given here:
%
\begin{eqnarray}
R^{\dagger}KR=&&
\left(\frac{h_{13}-h_{24}}{2}\sigma_{1}+
\frac{h_{14}+h_{23}}{2}\sigma_{3}\right)\otimes 1_{2}
+
1_{2}\otimes \left(\frac{h_{13}+h_{24}}{2}\sigma_{2}+
\frac{h_{14}-h_{23}}{2}\sigma_{3}\right)       \nonumber \\
&&+ \frac{h_{12}+h_{34}}{2}\sigma_{3}\otimes \sigma_{2}-
\frac{h_{12}-h_{34}}{2}\sigma_{1}\otimes \sigma_{3} \nonumber \\
&&
+\frac{h_{0}-h_{1}-h_{2}+h_{3}}{4}\sigma_{3}\otimes \sigma_{3}
+\frac{h_{0}+h_{1}-h_{2}-h_{3}}{4}\sigma_{1}\otimes \sigma_{2}
+\frac{h_{0}-h_{1}+h_{2}-h_{3}}{4}\sigma_{2}\otimes \sigma_{1}.
\nonumber \\
&&{}
\end{eqnarray}
%


\subsection{Approximate Result}

We do not know yet how to exponentiate the generic form of the ($R$-conjugated)
Hamiltonian (\ref{eq:adjoint-form}).
Our first main result is an approximate one for the generic case:
%
\begin{equation}
\label{eq:approximation}
U(t)\approx U_{1}(t)U_{2}(t)U_{3}(t)U_{4}(t),
\end{equation}
where each factor in  (\ref{eq:adjoint-form}) can be exponentiated exactly,
\begin{eqnarray}
\label{eq:special unitaries 1}
U_{1}(t)&=&R\left\{\exp\left(-it\left(\frac{h_{13}-h_{24}}{2}\sigma_{1}+
\frac{h_{14}+h_{23}}{2}\sigma_{3}\right)\right)\otimes 1_{2}
\right\}R^{\dagger},  \\
\label{eq:special unitaries 2}
U_{2}(t)&=&R\left\{1_{2}\otimes
\exp\left(-it\left(\frac{h_{13}+h_{24}}{2}\sigma_{2}+
\frac{h_{14}-h_{23}}{2}\sigma_{3}\right)\right)\right\}R^{\dagger},  \\
U_{3}(t)&=&
R\exp\left(-it\left(\frac{h_{12}+h_{34}}{2}\sigma_{3}\otimes \sigma_{2}
\right)\right)R^{\dagger},  \\
U_{4}(t)&=&
R\exp\left(-it\left(-\frac{h_{12}-h_{34}}{2}\sigma_{1}\otimes \sigma_{3}
\right)\right)R^{\dagger}.
\end{eqnarray}
As remarked earlier, $U_1$ and $U_2$ commute with each other and $U_3$ and $U_4$
also commute but the other pairs do not commute.
Hopefully the approximation
is not so bad. The non-exactness arises from the non-commutativity.

%

\subsection{Special Exact Result I}

If $h_{12}=h_{34}=0$ in (\ref{eq:adjoint-form}) the troublesome 
non-commutativity disappears.
Then we have the simple form
\begin{equation}
\label{eq:}
R^{\dagger}HR=
\left(\frac{h_{13}-h_{24}}{2}\sigma_{1}+
\frac{h_{14}+h_{23}}{2}\sigma_{3}\right)\otimes 1_{2}
+
1_{2}\otimes \left(\frac{h_{13}+h_{24}}{2}\sigma_{2}+
\frac{h_{14}-h_{23}}{2}\sigma_{3}\right)
\end{equation}
with
\begin{equation}
\label{eq:reduced-hamiltonian}
H=
\left(
   \begin{array}{cccc}
     0 & 0 & h_{13} & h_{14}  \\
     0 & 0 & h_{23} & h_{24}  \\
     h_{13} & h_{23} & 0 & 0  \\
     h_{14} & h_{24} & 0 & 0
   \end{array}
\right).
\end{equation}

Then we have the exact form
\begin{equation}
\label{eq:}
U(t)= U_{1}(t)U_{2}(t)= U_{2}(t)U_{1}(t)
\end{equation}
with $U_{1}(t)$ in (\ref{eq:special unitaries 1}) and $U_{2}(t)$ in
(\ref{eq:special unitaries 2}).

Though this Hamiltonian is of very special form there is some 
application shown in
the following section.

\subsection{Special Exact Result II}

The Hamiltonian that we really want to study is
\begin{equation}
\label{eq:more-reduced-hamiltonian}
H=
\left(
   \begin{array}{cccc}
     0 & h_{12} & 0 & h_{14}       \\
     h_{12} & 0 & h_{23} & 0  \\
     0 & h_{23} & 0 & h_{34}  \\
     h_{14} & 0 & h_{34} & 0
   \end{array}
\right),
\end{equation}
%
which is more general than the one discussed in recent papers by the author
  \cite{FHKW} \cite{KF1} and \cite{FF}, which is obtained by setting $h_{14}=0$.
  This restricted case was also discussed in  \cite{RZ}.
  At first sight this Hamiltonian looks rather different from 
(\ref{eq:reduced-hamiltonian}),
  which can be exactly exponentiated.
However, it can be reduced to the form of (\ref{eq:reduced-hamiltonian})
  by a similarity transformation in terms  of the exchange (swap) matrix $S$:
  \begin{equation}
\label{eq:SHS}
SHS=
\left(
   \begin{array}{cccc}
     0 & 0 & h_{12} & h_{14}        \\
     0 & 0 & h_{23} & h_{34}  \\
     h_{12} & h_{23} & 0 & 0  \\
     h_{14} & h_{34} & 0 & 0
   \end{array}
\right),
\end{equation}
with
\[
S=\frac{1}{2}(1_{2}\otimes 1_{2}+\sigma_{1}\otimes \sigma_{1}+
\sigma_{2}\otimes \sigma_{2}+\sigma_{3}\otimes \sigma_{3})=
\left(
   \begin{array}{cccc}
     1 & 0 & 0 & 0  \\
     0 & 0 & 1 & 0  \\
     0 & 1 & 0 & 0  \\
     0 & 0 & 0 & 1
   \end{array}
\right)\quad \Longrightarrow\quad S=S^{T}=S^{-1}.
\]
%
%
%
Therefore we have
\begin{equation}
\label{eq:}
R^{\dagger}SHSR=
\left(\frac{h_{12}-h_{34}}{2}\sigma_{1}+
\frac{h_{14}+h_{23}}{2}\sigma_{3}\right)\otimes 1_{2}
+
1_{2}\otimes \left(\frac{h_{12}+h_{34}}{2}\sigma_{2}+
\frac{h_{14}-h_{23}}{2}\sigma_{3}\right).
\end{equation}
%
The evolution operator
\[
U(t)=\mbox{e}^{-itH}
=SRR^{\dagger}S\mbox{e}^{-itH}SRR^{\dagger}S
=SR\mbox{e}^{-itR^{\dagger}SHSR}R^{\dagger}S
\]
takes an  exact factorized form
\begin{equation}
\label{eq:}
U(t)=U_{1}(t)U_{2}(t)=U_{2}(t)U_{1}(t)
\end{equation}
where
\begin{eqnarray}
U_{1}(t)&=&SR\left\{\exp\left(-it\left(\frac{h_{12}-h_{34}}{2}\sigma_{1}+
\frac{h_{14}+h_{23}}{2}\sigma_{3}\right)\right)\otimes 1_{2}
\right\}R^{\dagger}S,  \\
U_{2}(t)&=&SR\left\{1_{2}\otimes
\exp\left(-it\left(\frac{h_{12}+h_{34}}{2}\sigma_{2}+
\frac{h_{14}-h_{23}}{2}\sigma_{3}\right)\right)\right\}R^{\dagger}S.
\end{eqnarray}
%
Obviously the two factors commute and they belong to two independent $SU(2)$.
Compare this result with that of \cite{RZ}.

\section{B-C-H Formula for a class of matrices in SU(4)}

In this section we give a closed expression to the B-C-H formula for
the set of unitary matrices $\{\mbox{e}^{iH}\ |\ H\ \mbox{is a type of
(\ref{eq:more-reduced-hamiltonian})}\}$ by use of the results in the preceding
two sections. Let us prepare some notation for simplicity.

For two (real) symmetric matrices $A,\ B$
\begin{equation}
\label{eq:two matrices}
A=
\left(
   \begin{array}{cccc}
     0 & f_{1} & 0 & f_{4}      \\
     f_{1} & 0 & f_{2} & 0  \\
     0 & f_{2} & 0 & f_{3}  \\
     f_{4} & 0 & f_{3} & 0
   \end{array}
\right),\quad
B=
\left(
   \begin{array}{cccc}
     0 & g_{1} & 0 & g_{4}      \\
     g_{1} & 0 & g_{2} & 0  \\
     0 & g_{2} & 0 & g_{3}  \\
      g_{4} & 0 & g_{3} & 0
   \end{array}
\right)
\end{equation}
%
we can set
\[
R^{\dagger}(SAS)R={\bf a}_{1}\otimes 1_{2}+1_{2}\otimes {\bf a}_{2},
\quad
R^{\dagger}(SBS)R={\bf b}_{1}\otimes 1_{2}+1_{2}\otimes {\bf b}_{2}.
\]
Let us represent these four $su(2)$ elements in terms of the 
$\bf{R}^3$ vectors as
in (\ref{su2tor3}):
\begin{eqnarray*}
&&{\bf a}_{1}
    =\frac{f_{1}-f_{3}}{2}\sigma_{1}+
     \frac{f_{4}+f_{2}}{2}\sigma_{3},\quad
{\bf a}_{2}
    =\frac{f_{1}+f_{3}}{2}\sigma_{2}+
     \frac{f_{4}-f_{2}}{2}\sigma_{3}, \\
&&{\bf b}_{1}
    =\frac{g_{1}-g_{3}}{2}\sigma_{1}+
     \frac{g_{4}+g_{2}}{2}\sigma_{3},\quad
{\bf b}_{2}
    =\frac{g_{1}+g_{3}}{2}\sigma_{2}+
     \frac{g_{4}-g_{2}}{2}\sigma_{3}
\end{eqnarray*}
%
with
\[
{\bf \vec{a}}_{1}=
\left(
   \begin{array}{c}
   \frac{f_{1}-f_{3}}{2}  \\
   0                      \\
   \frac{f_{4}+f_{2}}{2}
   \end{array}
\right),
\quad
{\bf \vec{b}}_{1}=
\left(
   \begin{array}{c}
   \frac{g_{1}-g_{3}}{2}  \\
   0                      \\
   \frac{g_{4}+g_{2}}{2}
   \end{array}
\right);
\qquad
{\bf \vec{a}}_{2}=
\left(
   \begin{array}{c}
   0                     \\
   \frac{f_{1}+f_{3}}{2} \\
  \frac{f_{4}-f_{2}}{2}
   \end{array}
\right),
\quad
{\bf \vec{b}}_{2}=
\left(
   \begin{array}{c}
   0                       \\
   \frac{g_{1}+g_{3}}{2} \\
  \frac{g_{4}-g_{2}}{2}
   \end{array}
\right).
\]
%
Obviously ${\bf \vec{a}}_{1}$ and ${\bf \vec{b}}_{1}$  belong to the 
same $su(2)$,
whereas ${\bf \vec{a}}_{2}$ and ${\bf \vec{b}}_{2}$ belong to the other $su(2)$.
Next we introduce a pair of three real parameters 
$(\alpha_1,\beta_1,\gamma_1)$ and
$(\alpha_2,\beta_2,\gamma_2)$ defined for each pair of $\bf{R}^3$ vectors
$({\bf \vec{a}}_{1},{\bf \vec{b}}_{1})$ and $({\bf \vec{a}}_{2},{\bf 
\vec{b}}_{2})$:
\begin{eqnarray*}
&&\alpha_{1}=\alpha({\bf \vec{a}}_{1},{\bf \vec{b}}_{1}),\quad
\beta_{1}=\beta({\bf \vec{a}}_{1},{\bf \vec{b}}_{1}),\quad
\gamma_{1}=\gamma({\bf \vec{a}}_{1},{\bf \vec{b}}_{1}),    \\
&&\alpha_{2}=\alpha({\bf \vec{a}}_{2},{\bf \vec{b}}_{2}),\quad
\beta_{2}=\beta({\bf \vec{a}}_{2},{\bf \vec{b}}_{2}),\quad
\gamma_{2}=\gamma({\bf \vec{a}}_{2},{\bf \vec{b}}_{2}).
\end{eqnarray*}
See (\ref{alpadef}) in  the Appendix or the paper \cite{FS} for the 
definition of  these parameters. By combining the exact B-C-H formula 
for $su(2)$ for each pair  $({\bf \vec{a}}_{1},{\bf \vec{b}}_{1})$ 
and $({\bf \vec{a}}_{2},{\bf \vec{b}}_{2})$, we obtain
\begin{eqnarray}
\mbox{e}^{iA}\mbox{e}^{iB}
&=&
SRR^{\dagger}S\mbox{e}^{iA}SRR^{\dagger}S\mbox{e}^{iB}SRR^{\dagger}S
\nonumber \\
&=&
SR\mbox{e}^{iR^{\dagger}SASR}\mbox{e}^{iR^{\dagger}SBSR}R^{\dagger}S
\nonumber \\
&=&
SR
\mbox{e}^{i({\bf a}_{1}\otimes 1_{2}+1_{2}\otimes {\bf a}_{2})}
\mbox{e}^{i({\bf b}_{1}\otimes 1_{2}+1_{2}\otimes {\bf b}_{2})}
R^{\dagger}S  \nonumber \\
&=&
SR
\left(\mbox{e}^{i{\bf a}_{1}}\otimes \mbox{e}^{i{\bf a}_{2}}\right)
\left(\mbox{e}^{i{\bf b}_{1}}\otimes \mbox{e}^{i{\bf b}_{2}}\right)
R^{\dagger}S  \nonumber \\
&=&
SR
\left(\mbox{e}^{i{\bf a}_{1}}\mbox{e}^{i{\bf b}_{1}}\right)\otimes
\left(\mbox{e}^{i{\bf a}_{2}}\mbox{e}^{i{\bf b}_{2}}\right)
R^{\dagger}S  \nonumber \\
&=&
SR
\mbox{e}^{i(\alpha_{1}{\bf a}_{1}+\beta_{1}{\bf b}_{1}+
\gamma_{1}\frac{i}{2}[{\bf a}_{1},{\bf b}_{1}])}
\otimes
\mbox{e}^{i(\alpha_{2}{\bf a}_{2}+\beta_{2}{\bf b}_{2}+
\gamma_{2}\frac{i}{2}[{\bf a}_{2},{\bf b}_{2}])}
R^{\dagger}S  \nonumber \\
&=&
SR
\mbox{e}^{i\left\{
(\alpha_{1}{\bf a}_{1}+\beta_{1}{\bf b}_{1}+
\gamma_{1}\frac{i}{2}[{\bf a}_{1},{\bf b}_{1}])\otimes 1_{2}
+
1_{2}\otimes (\alpha_{2}{\bf a}_{2}+\beta_{2}{\bf b}_{2}+
\gamma_{2}\frac{i}{2}[{\bf a}_{2},{\bf b}_{2}])
\right\}}
R^{\dagger}S  \nonumber \\
&=&
\mbox{e}^{iSR\left\{
(\alpha_{1}{\bf a}_{1}+\beta_{1}{\bf b}_{1}+
\gamma_{1}\frac{i}{2}[{\bf a}_{1},{\bf b}_{1}])\otimes 1_{2}
+
1_{2}\otimes (\alpha_{2}{\bf a}_{2}+\beta_{2}{\bf b}_{2}+
\gamma_{2}\frac{i}{2}[{\bf a}_{2},{\bf b}_{2}])
\right\}R^{\dagger}S} \nonumber \\
&\equiv&
\mbox{e}^{i{BCH}(A,B)}.
\end{eqnarray}
The desired closed form of the B-C-H formula reads
\begin{eqnarray}
\label{eq:B-C-H closed}
&&{BCH}(A,B) \nonumber \\
&=&SR\left\{
\left(\alpha_{1}{\bf a}_{1}+\beta_{1}{\bf b}_{1}+
\gamma_{1}\frac{i}{2}[{\bf a}_{1},{\bf b}_{1}]\right)\otimes 1_{2}
+
1_{2}\otimes \left(\alpha_{2}{\bf a}_{2}+\beta_{2}{\bf b}_{2}+
\gamma_{2}\frac{i}{2}[{\bf a}_{2},{\bf b}_{2}]\right)
\right\}R^{\dagger}S   \nonumber \\
&=&
\left(
   \begin{array}{cccc}
      0 & (12) & (13) & (14)                                  \\
     \overline{(12)} & 0 & (23) & (24)                        \\
     \overline{(13)} & \overline{(23)} & 0 & (34)             \\
     \overline{(14)} & \overline{(24)} & \overline{(34)} & 0
   \end{array}
\right),
\end{eqnarray}
whose entries are
\begin{eqnarray*}
(12)&=&
\alpha_{1}\frac{f_{1}-f_{3}}{2}+\beta_{1}\frac{g_{1}-g_{3}}{2}+
\alpha_{2}\frac{f_{1}+f_{3}}{2}+\beta_{2}\frac{g_{1}+g_{3}}{2}, \\
(13)&=&
i\left\{
\gamma_{1}\left(\frac{f_{1}-f_{3}}{2}\frac{g_{4}+g_{2}}{2}-
\frac{f_{4}+f_{2}}{2}\frac{g_{1}-g_{3}}{2}\right)-
\gamma_{2}\left(\frac{f_{1}+f_{3}}{2}\frac{g_{4}-g_{2}}{2}-
\frac{f_{4}-f_{2}}{2}\frac{g_{1}+g_{3}}{2}\right)
\right\}, \\
(14)&=&
\alpha_{1}\frac{f_{4}+f_{2}}{2}+\beta_{1}\frac{g_{4}+g_{2}}{2}+
\alpha_{2}\frac{f_{4}-f_{2}}{2}+\beta_{2}\frac{g_{4}-g_{2}}{2}, \\
(23)&=&
\alpha_{1}\frac{f_{4}+f_{2}}{2}+\beta_{1}\frac{g_{4}+g_{2}}{2}-
\alpha_{2}\frac{f_{4}-f_{2}}{2}-\beta_{2}\frac{g_{4}-g_{2}}{2}, \\
(24)&=&
i\left\{
\gamma_{1}\left(\frac{f_{1}-f_{3}}{2}\frac{g_{4}+g_{2}}{2}-
\frac{f_{4}+f_{2}}{2}\frac{g_{1}-g_{3}}{2}\right)+
\gamma_{2}\left(\frac{f_{1}+f_{3}}{2}\frac{g_{4}-g_{2}}{2}-
\frac{f_{4}-f_{2}}{2}\frac{g_{1}+g_{3}}{2}\right)
\right\}, \\
(34)&=&
-\alpha_{1}\frac{f_{1}-f_{3}}{2}-\beta_{1}\frac{g_{1}-g_{3}}{2}+
\alpha_{2}\frac{f_{1}+f_{3}}{2}+\beta_{2}\frac{g_{1}+g_{3}}{2}.
\end{eqnarray*}
%
Here $ \overline{(12)}$ is the complex conjugate of $(12)$.

This is another main result of the present paper.

\section{Discussion}

In this letter we addressed the problem of explicit exponentiation of 
the generic Hamiltonian
  in the general four level system.
  With the aid of the magic matrix, we obtained an approximate result 
for the generic case and some exact results for certain restricted 
forms of the Hamiltonian.
  It is a good challenge to derive the explicit form of the evolution 
operator for the most generic Hamiltonian in the four level system.





\vspace{10mm}
\noindent{\em Acknowledgments}\\
The author wishes to thank K. Funahashi, H. Oike, R. Sasaki and T. Suzuki for 
their helpful comments and suggestions.

\section*{Appendix:\ \ B-C-H Formula for SU(2)}
%
In this Appendix we recapitulate the closed form expression of the 
B-C-H formula for
$SU(2)$ in the $2\times2$ representation, see \cite{FS} for more details.

First of all let us recall the well-known exponentiation formula:
\begin{eqnarray}
\mbox{e}^{i(x\sigma_{1}+y\sigma_{2}+z\sigma_{3})}
&=&
\cos{r}1_{2}+\frac{\sin{r}}{r}i(x\sigma_{1}+y\sigma_{2}+z\sigma_{3}) \\
&=&
\left(
   \begin{array}{cc}
     \cos{r}+i\frac{\sin{r}}{r}z & i\frac{\sin{r}}{r}(x-iy) \\
     i\frac{\sin{r}}{r}(x+iy) & \cos{r}-i\frac{\sin{r}}{r}z
   \end{array}
\right),
\end{eqnarray}
where $r=\sqrt{x^{2}+y^{2}+z^{2}}$. This is a simple exercise.

For the group $SU(2)$ it is easy to sum up all terms in the B-C-H expansion
by  using  the above exact exponentiation formula.
Because of the obvious relation $su(2)\cong so(3)$, three dimensional 
vector notation is quite useful for $su(2)$.
For  $2\times2$ hermitian matrices, %
\[
X=x_{1}\sigma_{1}+x_{2}\sigma_{2}+x_{3}\sigma_{3},\quad
Y=y_{1}\sigma_{1}+y_{2}\sigma_{2}+y_{3}\sigma_{3}\in H_{0}(2,\fukuso)
\]
we associate $\bf{R}^3$ vectors.
  Namely, we set
\begin{equation}
X\ \longrightarrow\ {\bf x}=
\left(
   \begin{array}{c}
     x_{1} \\
     x_{2} \\
     x_{3}
   \end{array}
\right),\quad
Y\ \longrightarrow\ {\bf y}=
\left(
   \begin{array}{c}
     y_{1} \\
     y_{2} \\
     y_{3}
   \end{array}
\right)
\label{su2tor3}
\end{equation}
%
%
and
\[
{\bf x}\cdot {\bf y}=x_{1}y_{1}+x_{2}y_{2}+x_{3}y_{3}=\mbox{Tr}(XY)/2,\quad
|{\bf x}|=\sqrt{{\bf x}\cdot {\bf x}},\quad
|{\bf y}|=\sqrt{{\bf y}\cdot {\bf y}}.
\]
Then the commutator of $X$ and $Y$ corresponds to the vector product
${\bf x}\times {\bf y}$:
\[
-\frac{i}{2}[X,Y]\ \longrightarrow\
{\bf x}\times {\bf y}=
\left(
   \begin{array}{c}
     x_{2}y_{3}-x_{3}y_{2} \\
     x_{3}y_{1}-x_{1}y_{3} \\
     x_{1}y_{2}-x_{2}y_{1}
   \end{array}
\right).
\]

\par \vspace{5mm}
Now we are in a position to state the B-C-H formula for $SU(2)$:
\begin{equation}
\mbox{e}^{iX}\mbox{e}^{iY}=\mbox{e}^{iZ}\ :\quad
Z=\alpha X + \beta Y + \gamma \frac{i}{2}[X,Y],
\end{equation}
where the real coefficients $\alpha$,  $\beta$ and $\gamma$ are defined by
\begin{eqnarray}
&&\alpha\equiv \alpha({\bf x},{\bf y})=
\frac{\sin^{-1}\rho}{\rho}
\frac{\sin{|{\bf x}|}\cos{|{\bf y}|}}{|{\bf x}|},\quad
\beta\equiv \beta({\bf x},{\bf y})=
\frac{\sin^{-1}\rho}{\rho}
\frac{\cos{|{\bf x}|}\sin{|{\bf y}|}}{|{\bf y}|},  \nonumber \\
&&\gamma\equiv \gamma({\bf x},{\bf y})=
\frac{\sin^{-1}\rho}{\rho}
\frac{\sin{|{\bf x}|}\sin{|{\bf y}|}}{|{\bf x}||{\bf y}|}
\label{alpadef}
\end{eqnarray}
with
\begin{eqnarray*}
\rho^{2}&\equiv& \rho({\bf x},{\bf y})^{2}  \\
&=&
\sin^{2}{|{\bf x}|}\cos^{2}{|{\bf y}|}+\sin^{2}{|{\bf y}|}\\
&&\qquad\qquad
-
\frac{\sin^{2}{|{\bf x}|}\sin^{2}{|{\bf y}|}}{|{\bf x}|^{2}|{\bf y}|^{2}}
\left({\bf x}\cdot {\bf y}\right)^{2}
+
\frac{2\sin{|{\bf x}|}\cos{|{\bf x}|}
\sin{|{\bf y}|}\cos{|{\bf y}|}}{|{\bf x}||{\bf y}|}
\left({\bf x}\cdot {\bf y}\right).
\end{eqnarray*}

%
The proof is not difficult, so it is left to the readers.



\begin{thebibliography}{99}
%
\bibitem{FHKW}K. Fujii, K. Higashida, R. Kato and Y. Wada :
\newblock A Rabi Oscillation in Four and Five Level Systems,
\newblock Yokohama Mathematical Journal, {\bf 53} (2006), 63,
\newblock quant-ph/0312060.
%
\bibitem{KF1}K. Fujii :
\newblock Study on Dynamics of N Level System of Atom by Laser Fields,
\newblock Contemporary Mathematics and Its Applications (in Russian), 
{\bf 44} (2007), 3, 
\newblock quant-ph/0512126.
%
\bibitem{FF}K. Fujii :
\newblock A Generalization of the Preceding Paper ``A Rabi Oscillation 
in Four and Five Level Systems",
\newblock quant-ph/0605132.
%
\bibitem{YMa}Y. Makhlin :
\newblock Nonlocal properties of two-qubit gates and mixed states
and optimization of quantum computations,
\newblock Quant. Info. Proc. {\bf 1} (2002), 243,
\newblock quant-ph/0002045.
%
\bibitem{FOS}K. Fujii, H. Oike and T. Suzuki :
\newblock More on the isomorphism $SU(2)\otimes SU(2)\cong SO(4)$,
\newblock Int. J. Geom. Methods Mod. Phys. {\bf 4} (2007), 471, 
\newblock quant-ph/0608186.
%
\bibitem{ZVWS}J. Zhang, J. Vala, K. B. Whaley and S. Sastry :
\newblock A geometric theory of non-local two-qubit operations,
\newblock Phys. Rev. A {\bf 67} (2003), 042313,
\newblock quant-ph/0209120.
%
\bibitem{RZ}V. Ramakrishna and H. Zhou :
\newblock On The Exponential of Matrices in $su(4)$,
\newblock math-ph/0508018.
%
\bibitem{Vara}V. S. Varadarajan :
\newblock Lie Groups, Lie Algebras, and Their Representations,
\newblock New York, Springer 1984.
%
\bibitem{YS}T. Yamanouchi and M. Sugiura :
\newblock Introduction to Topological Groups (in Japanese),
\newblock Tokyo, Baifukan 1960.
%
\bibitem{SW}S. Weigert :
\newblock Baker-Campbell-Hausdorff relation for special unitary groups
$SU(N)$,
\newblock J. Phys. A {\bf 30} (1997), 8739,
\newblock quant-ph/9710024.
%
\bibitem{FS}K. Fujii and T. Suzuki :
\newblock On the Magic Matrix by Makhlin and the B-C-H Formula in $SO(4)$,
\newblock to appear in Int. J. Geom. Methods Mod. Phys. {\bf 4} (2007) no.6,
\newblock quant-ph/0610009.
%
\bibitem{KF3}K. Fujii :
\newblock Introduction to Coherent States and Quantum Information Theory,
\newblock quant-ph/0112090.
%
\bibitem{KF2}K. Fujii :
\newblock Introduction to Grassmann Manifolds and Quantum Computation,
\newblock J. Applied Math, {\bf 2} (2002), 371,
\newblock quant-ph/0103011.
%
\end{thebibliography}
\end{document}